\documentclass[10pt,fleqn]{article}

\usepackage{amssymb}

\newcommand{\name}[1]{\begin{flushleft}
                       \LARGE \bf #1
                       \end{flushleft}\vspace{-3mm}}

\newcommand{\Author}[1]{\begin{flushleft}
                       \it #1 \end{flushleft}}

\newcommand{\Adress}[1]{\begin{flushleft}
                       \it #1 \end{flushleft}}

\newcommand{\be}{\begin{equation}}
\newcommand{\ee}{\end{equation}}
\newcommand{\ba}{\hspace*{-5pt}\begin{array}}
\newcommand{\ea}{\end{array}}

\newcommand{\pbf}[1]{\mbox{\mathversion{bold}$#1$}}

\begin{document}

\name{On the $\pbf{CP}$-noninvariant equations for \\
the particle with zero mass and spin
$\pbf{s=\frac 12}$}

\medskip

\noindent{published in {\it Lettere al Nuovo Cimento}, 1970,  {\bf 4}, N~20, P. 927--928.}

\Author{Wilhelm I. FUSHCHYCH and A.L. GRISHCHENKO}

\Adress{Institute of Mathematics of the National Academy of
Sciences of Ukraine, \\ 3 Tereshchenkivska  Street, 01601 Kyiv-4,
UKRAINE}

\noindent {\tt URL:
http://www.imath.kiev.ua/\~{}appmath/wif.html\\ E-mail:
symmetry@imath.kiev.ua}

\medskip

\noindent
One of us [1] has shown that for the particle with zero mass and spin $s=\frac 12$
there are three types of two-component equations (or one four-component equation with three
different subsidiary conditions) which differ from one another by $P$, $T$ and $C$
properties. One of these equations is the two-component Weyl equation which, as is well known,
is equivalent to the four-component Dirac equation
\be
\gamma_\mu p^\mu \Psi(t,\pbf{x})=0, \qquad \mu=0,1,2,3,
\ee
with the subsidiary relativistic invariant condition
\be
(1+\gamma_5) \Psi(t,\pbf{x})=0.
\ee
Equations (1), (2) may be written in the form of a single equation~[2]
\be
\left\{ \gamma_\mu p^\mu +\varkappa_1 (1+\gamma_5)\right\} \Psi(t,\pbf{x})=0,
\ee
where $\varkappa_1$ is an arbitrary constant (not connected with mass of the particle.
Equation (3) (or eqs.~(1) and~(2)) is  $P$ and $C$ noninvariant, but $CP$-invariant.

In this note we give two other relativistic invariant equations which differ from~(3) (or from~(1)
with the subsidiary condition~(2)).

These equations have the form
\be
\left\{ \gamma_\mu p_\mu +\varkappa_2 \left(1+\gamma_5 \frac{H}{E}\right)\right\}\Psi(t,\pbf{x})=0,
\ee
\be
\left\{ \gamma_\mu p^\mu +\varkappa_3 \left(1+ \frac{H}{E}\right)\right\}\Psi(t,\pbf{x})=0,
\ee
\be
H=\gamma_0 \gamma_k p_k, \qquad k=1,2,3, \qquad E=\sqrt{p_1^2 +p_2^2 +p_3^2},
\ee
$\varkappa_2$, $\varkappa_3$ are arbitrary constants.

Equation (4) is equivalent to eq. (1) with the subsidiary condition
\be
\left(1+\gamma_5 \frac{H}{E}\right) \Psi(t,\pbf{x})=0.
\ee

Equation (5) is equivalent to eq. (1) with the subsidiary condition
\be
\left(1+\frac{H}{E}\right) \Psi(t,\pbf{x})=0.
\ee

The relativistic invariance of eqs. (4) and (5) (or the invariance of the subsidiary conditions~(7)
and~(8)) follows from the fact that the operators $\gamma_5$ and $H/E$ are invariants of
the Poincar\'e group (for the case of zero mass).

It easy to verify that eq. (4) (or eq. (1) with condition (7)) is $CP$ and $CPT$ noninvariant.

Equation (5) (or eq. (1) with condition (8)) is $P$ and $T$ invariant (in the sense of Wigner time
reflection), but $C$-noninvariant.

Equation (4) coincides with the equation obtained earlier~[1] (where the sub\-sti\-tu\-tion
$e_3 =p_3/|p_3|\to H/E$ should be made).

Thus, as distinguished from eq. (3) (eqs.~(1) and~(2)) there are two more eqs.~(4) and~(5) which
are also relativistic invariant, but $CP$-noninvariant.

A more detailed analysis of eqs. (4) and (5) will be given in another paper.

\medskip

\begin{enumerate}

\footnotesize

\item Fushchych W.I., {\it Nucl. Phys. B}, 1970, {\bf  21}, 321.\ \ {\tt quant-ph/0206077}

\item Tokuoka Z.,  {\it Progr. Theor. Phys.}, 1967, {\bf 37}, 603;\\
Sen Gupta N.D., {\it Nucl. Phys. B}, 1968, {\bf  4}, 147.

\end{enumerate}
\end{document}